\documentclass[showpacs,aps,prb,twocolumn,superscriptaddress,preprintnumbers,amsmath,amssymb,floatfix,10pt]{revtex4}
\usepackage{graphicx}
\usepackage{dcolumn}
\usepackage{float}
\usepackage{bm}
\usepackage{enumerate}
\usepackage[english]{babel}
\usepackage{amsfonts}
\usepackage{amsthm}
\usepackage{amssymb}
\usepackage{calc}
\usepackage{color}
\usepackage{ifthen}
\usepackage{hyperref}
\graphicspath{{}{figures/}}
\newcommand{\bfw}{\textbf{w}}
\newcommand{\bfz}{\textbf{z}}

\newcommand{\bfr}{\textbf{r}}

\newcommand{\bfq}{\textbf{q}}
\newcommand{\bfG}{\textbf{G}}

\newcommand{\im}{\mathrm{i}}             
\usepackage{inputenc}
\inputencoding{utf8}

\begin{document}
\title{Freezing of an unconventional two-dimensional plasma}

\author{Egil V. Herland}
\affiliation{Department of Physics, Norwegian University of Science and Technology, N-7491 Trondheim, Norway}

\author{Egor Babaev}
\affiliation{Physics Department, University of Massachusetts, Amherst, Massachusetts 01003, USA}
\affiliation{Department of Theoretical Physics, The Royal Institute of Technology, 10691 Stockholm, Sweden}

\author{Parsa Bonderson}
\affiliation{Station Q, Microsoft Research, Santa Barbara, California 93106-6105, USA}

\author{Victor Gurarie}
\affiliation{Department of Physics, CB 390, University of Colorado, Boulder, Colorado 80309, USA}

\author{Chetan Nayak}
\affiliation{Station Q, Microsoft Research, Santa Barbara, California 93106-6105, USA}
\affiliation{Department of Physics, University of California, Santa Barbara, California 93106, USA}

\author{Leo Radzihovsky}
\affiliation{Department of Physics, CB 390, University of Colorado, Boulder, Colorado 80309, USA}

\author{Asle Sudb\o}
\affiliation{Department of Physics, Norwegian University of Science and Technology, N-7491 Trondheim, Norway}

\begin{abstract}
We study an unconventional two-dimensional, two-component classical plasma on a sphere, with emphasis on detecting signatures 
of melting transitions. This system is relevant to Ising-type quantum Hall states, and is unconventional in the sense that it features
particles interacting via two different two-dimensional Coulomb interactions. One species of particles in the plasma carries charge 
of both types $({Q_1},{Q_2})$, while the other species carries only charge of the second type $({0},-{Q_2})$. We find signatures of a 
freezing transition at $Q_1^2 \simeq 140$. This means that the species with charge of both types will form a Wigner crystal, 
whereas the species with charge of the second type also shows signatures of being a Wigner crystal, due to the attractive 
inter-component interaction of the second type. Moreover, there is also a Berezinskii-Kosterlitz-Thouless phase transition 
at $Q_2^2  \simeq 4$, at which the two species of particles bind to form molecules that are 
neutral with respect to the second Coulomb interaction. These two transitions appear to be independent of each other, giving 
a rectangular phase diagram. As a special case, $Q_2=0$ describes the (conventional) two-dimensional one-component plasma. Our study 
is consistent with previous studies of this plasma, and sheds new light on the freezing transition of this system. 
\end{abstract}


\maketitle

\section{Introduction}
Multi-component quantum condensates with novel types of inter-component interactions are of considerable interest in contemporary
physics. For example, they are relevant to widely disparate systems, including low-dimensional spin-$1/2$ quantum 
antiferromagnets~\cite{Sachdev_MPLB_1990, Senthil_Science_2004, Senthil_PRB_2004,Kuklov_AnnPhys_2006,Kragset_PRL_2006}, Bose-Einstein 
condensates~\cite{Dahl_PRBa_2008, Dahl_PRBb_2008, Dahl_PRL_2008},
multi-component/multi-band superconductors~\cite{Babaev_Nat_2004, Babaev_Nat_Phys_2007, Smiseth_PRB_2005}, and non-Abelian quantum Hall states 
and topological superconductors~\cite{Bonderson_PRB_2011}. These systems have the remarkable property of possessing a mapping to a classical 
multi-component plasma system with highly unusual intra- and inter-component interactions. The statistical properties of these 
unconventional plasmas -- especially their phase diagrams -- have important ramifications for the physics of their corresponding 
fractional quantum Hall systems~\cite{Bonderson_PRB_2011}. The statistical physics of such systems has only recently begun to be explored. 
In a previous paper, we investigated the metal-insulator transition in a particular version of such a plasma~\cite{Herland_PRB_2012}. In this 
paper, we will extend these investigations to a study of the freezing of such a plasma from a liquid to a Wigner crystal.

The canonical partition function of the unconventional two-component plasma that we investigate is given by~\cite{Bonderson_PRB_2011, Herland_PRB_2012}
\begin{equation}
Z = \int \left(\prod_{i=1}^N \textrm{d}^2z_i \right) \left(\prod_{a=1}^N \textrm{d}^2w_a \right)\mathrm{e}^{-V},
\label{Eq:Model_Z}
\end{equation}
where the potential energy
\begin{align}
V = & \quad -Q_2^2 \sum_{a<b=1}^N \ln|\bfw_a-\bfw_b| + Q_2^2 \sum_{a,i=1}^N \ln|\bfz_i-\bfw_a|\nonumber\\
& - (Q_1^2 + Q_2^2)\sum_{i<j=1}^N \ln|\bfz_i-\bfz_j| + V_{z, \textnormal{BG}}
\label{Eq:Model_S}
\end{align}
describes two species (components) of particles interacting via two different types of two-dimensional (2D) Coulomb interactions, which are logarithmic.
Here, the $\bfz_i$ are coordinate vectors for the $N$ particles of component $z$, which carry charge $Q_1$ of the first interaction (type 1) and charge $Q_2$ 
of the second interaction (type 2). The $\bfw_a$ are coordinate vectors for the $N$ particles of component $w$, which carry no charge of type 1 and charge 
$-Q_2$ of type 2. The term $V_{z, \textnormal{BG}}$ describes the interaction of the $z$-particles with a uniform density neutralizing background charge. Note that the form given in Eq.~\eqref{Eq:Model_Z} implies that the temperature $T = 1$.

This plasma is related~\cite{Bonderson_PRB_2011} to inner products of quantum-mechanical trial wave functions of Ising-type quantum
Hall states, such as the Moore-Read Pfaffian~\cite{Moore_Read_NPB_1991}, anti-Pfaffian~\cite{Lee_PRL_2007,Levin_PRL_2007}, and Bonderson-Slingerland 
hierarchy states~\cite{Bonderson_PRB_2008}. For the charge values relevant to these states, the plasma was shown to be in its metallic liquid 
phase~\cite{Herland_PRB_2012}, which allows for the calculation of the braiding statistics of quasiparticle excitations of these 
states~\cite{Bonderson_PRB_2011}, confirming their conjectured non-Abelian statistics. This plasma is also related to rotating two-component 
Bose-Einstein condensates (BECs) in two dimensions~\cite{Herland_PRB_2012}.

While Ref.~\onlinecite{Herland_PRB_2012} focused on the cases $Q_1 = 0, 2$, which are particularly relevant for Ising-type quantum Hall states, here we 
will investigate the plasma for large values of $Q_1$. In the limit in which $Q_2 = 0$, the $w$-particles do not interact, and the plasma thus reduces to 
the standard 2D one-component Coulomb plasma (OCP). It is generally believed that, at high values of $Q_1$, the OCP will be in a 2D solid state in which
the charges form a triangular lattice with quasi-long-range translational and long-range orientational order \cite{DSFisher_1982}, as found in the simulations in 
Refs.~\onlinecite{de_Leeuw_PhysicaA_1982, Caillol_JStatPhys_1982, Choquard_PRL_1983, Franz_PRL_1994}. However, some 
studies have claimed that there is no low-temperature (high-$Q_1$) crystalline state in the OCP, due to
the proliferation of screened disclinations~\cite{O'Neill_PRB_1993, Dodgson_PRB_1997, Moore_PRL_1999, McClarty_PRB_2007}.

If we assume that the generally-held view is correct (and we present evidence supporting this view), so that there is a low-temperature crystalline 
state, then the melting of this crystal can occur according to either of two possible scenarios. One possibility is the Kosterlitz-Thouless-Halperin-Nelson-Young 
theory (KTHNY)~\cite{Kosterlitz_JPhysC_1973, Halperin_PRL_1978,Nelson_PRB_1979, Young_PRB_1979}, according to which dislocation pairs 
unbind via a vector-defects version of the BKT transition at a
Berezinskii-Kosterlitz-Thouless (BKT) like transition. The system then enters a hexatic liquid phase in which there is no translational order, but there is 
quasi-long-range hexatic order. Then, there is a second BKT transition at which disclination pairs unbind, hexatic order is
lost, and the system enters an isotropic liquid phase. The other possibility is a direct first-order melting transition at a lower temperature
than the KTHNY-theory predicts~\cite{Halperin_PRL_1978}. There have been considerable efforts to investigate 2D melting, both experimentally and
by numerical simulations. Some studies have found KTHNY transitions while others have found a weakly first-order melting 
transition~\cite{Strandburg_RMP_1988, Chen_PRL_1995, Dash_RMP_1999, Zahn_PRL_1999, Bates_PRE_2000, Dietel_PRB_2006,Lee_PRE_2008, Guillamon_NatPhys_2009}. 
It appears that the nature of 2D melting depends on details of the interatomic potential. In the case of logarithmic interactions, most numerical simulations
find a first-order transition~\cite{de_Leeuw_PhysicaA_1982,Caillol_JStatPhys_1982, Choquard_PRL_1983, Franz_PRL_1994}.

Before proceeding to a description of our simulations, we mention that, in principle, there is one other possibility: a Lifshitz transition 
from the liquid to a striped or ``microemulsion'' phase and then later to a Wigner crystal, as discussed by Kivelson and 
Spivak~\cite{Kivelson-Spivak}. Such a scenario must be considered when there is a linear coupling between the order parameter and the 
uniform density (i.e., between the order parameter at wavevector ${\bf q}$ and the density at wavevector $-{\bf q}$) or, equivalently,
when the first derivative of the energy with respect to the density is discontinuous at the transition. However, in our case, the order 
parameter is the density at non-zero wavevector, so no such  linear coupling can occur. Furthermore, the order parameter vanishes on both 
sides of the transition since the crystalline phase is only quasi-long-range ordered, so there would be no discontinuity even if there 
were a linear coupling. However, even in systems to which the Kivelson-Spivak~\cite{Kivelson-Spivak} argument applies, there are two possible
scenarios, similar to the ones that we consider: a direct first-order phase transition (which is permitted for the case of logarithmic
interactions) and a continuous transition via one or more intermediate phases.

\section{Model and Simulation}

The system described in Eqs.~\eqref{Eq:Model_Z} and~\eqref{Eq:Model_S} is studied by means of large-scale Monte Carlo simulations
on a sphere of radius $R$. In this geometry, the distance between two points ${\bf r}_1$ and ${\bf r}_2$ is taken to be the chord length
\begin{equation}
|{\bf r}_1-{\bf r}_2| = \sqrt{2} R \, (1- \hat{\bf r}_1 \cdot \hat{\bf r}_2)^{\frac{1}{2}},
\end{equation}
and the term $V_{z, \textnormal{BG}}$ is simply a uniform constant that can be disregarded.
Hence, the model in Eq.~\eqref{Eq:Model_S} may be written in the form (up to constant terms)
\cite{Herland_PRB_2012, Caillol_JStatPhys_1982, Caillol_PRB_1986, Caillol_JChemPhys_1991}
\begin{align}
  V = \frac{1}{2}\Bigg [& Q_2^2 \sum_{a,i=1}^N \ln(1-\hat{\bfz}_i\cdot\hat{\bfw}_a)-Q_2^2 \sum_{a<b=1}^N \ln(1-\hat{\bfw}_a\cdot
  \hat{\bfw}_b)\nonumber\\
  &- (Q_1^2+Q_2^2)\sum_{i<j=1}^N \ln(1-\hat{\bfz}_i\cdot\hat{\bfz}_j)\Bigg].
  \label{Eq:Model_S_sphere}
\end{align}
Here, $\hat{\bfw}_a$, $\hat{\bfz}_i$ are the positions of the particles on the surface of the unit sphere. Details of the derivation, as well
as on the technicalities of the Monte Carlo simulations, are presented in Ref.~\onlinecite{Herland_PRB_2012}. Moreover, to improve sampling
at high values of $Q_1$, we used the parallel tempering algorithm~\cite{Hukushima_JPSJ_96, Earl_PCCP_05}, where the set of couplings was
found by measuring first-passage-times as described in Ref.~\onlinecite{Nadler_PRE_08}.

In addition to the logarithmic interactions, we regularize the attractive interactions by adding a short-range hard-core repulsion such that 
particles are not permitted to be closer than the particle diameter $d$. Hence, there is a nonzero dimensionless density $\eta = 2Ns/A$ where 
$s = \pi d^2/4$ and $A$ is the area of the system.

\section{Results for the One-Component Plasma}

First, we consider the case in which $Q_2 = 0$. This is motivated by the fact that previous studies of the OCP on the surface of a sphere are 
not consistent. In Ref.~\onlinecite{Caillol_JStatPhys_1982}, a freezing transition at $Q_1^2 \simeq 140$ was found by comparing the free energy 
of the solid and liquid state. However, in Ref.~\onlinecite{Moore_PRL_1999}, the absence of a finite-temperature crystalline state was claimed 
and numerical evidence supporting this was provided, essentially by showing that the correlation length for crystalline order was 
non-divergent, $\xi \propto Q_1$ for all $Q_1^{-1} > 0$. We will return to this below.

The structure function is given by
\begin{eqnarray}
S({\bf q}) & \equiv & \frac{1}{N} \int \int d {\bf{r}} ~ d {\bf{r}}^{\prime} \mathrm{e}^{\im {\bf q} \cdot ( {\bf{r}} -   {\bf{r}}^{\prime})} 
\langle  n({\bf r})  n({\bf r}^{\prime})\rangle \nonumber \\
& = & 1 + \frac{1}{N} \sum_{i \neq j} \langle \mathrm{e}^{\im {\bf q} \cdot ( {\bf{r}}_i -   {\bf{r}}_j  )} \rangle.
\label{Eq:Structure_function}
\end{eqnarray}
Here, $\langle  n({\bf r})  n({\bf r}^{\prime})\rangle$ 
is the density-density function, where $n({\bf r}) = \sum_j \delta({\bf{r}} - {\bf{r}}_j)$ is the local density, and $\langle \dots \rangle$ 
denotes a statistical average. Furthermore, $\{ {\bf{r}}_j \}$ denotes the positions of the point particles in the problem and $\bfq$ is the Fourier space vector. In this work, we measure the azimuthal average of $S({\bf q})$ modified for a spherical geometry, given by~\cite{Caillol_JStatPhys_1982, Moore_PRL_1999, Hansen_PRL_1979}
\begin{equation}
S(q) = 1 + 2\pi n R^2 \int_{0}^{\pi} \textrm{d}\theta \left[g(R\theta)-1\right]
\sin\theta J_0(qR\theta),
\label{Eq:Structure_factor}
\end{equation}
where $n$ is the number density, $R$ is the radius of the sphere on which the particles live, $g(R\theta)$ is the pair distribution function
with $\theta$ as the chord angle, $J_0(x)$ is a zeroth order Bessel function, and $q$ is the magnitude of $\bfq$. 

\begin{figure}[tbp]
  \includegraphics[width=\columnwidth]{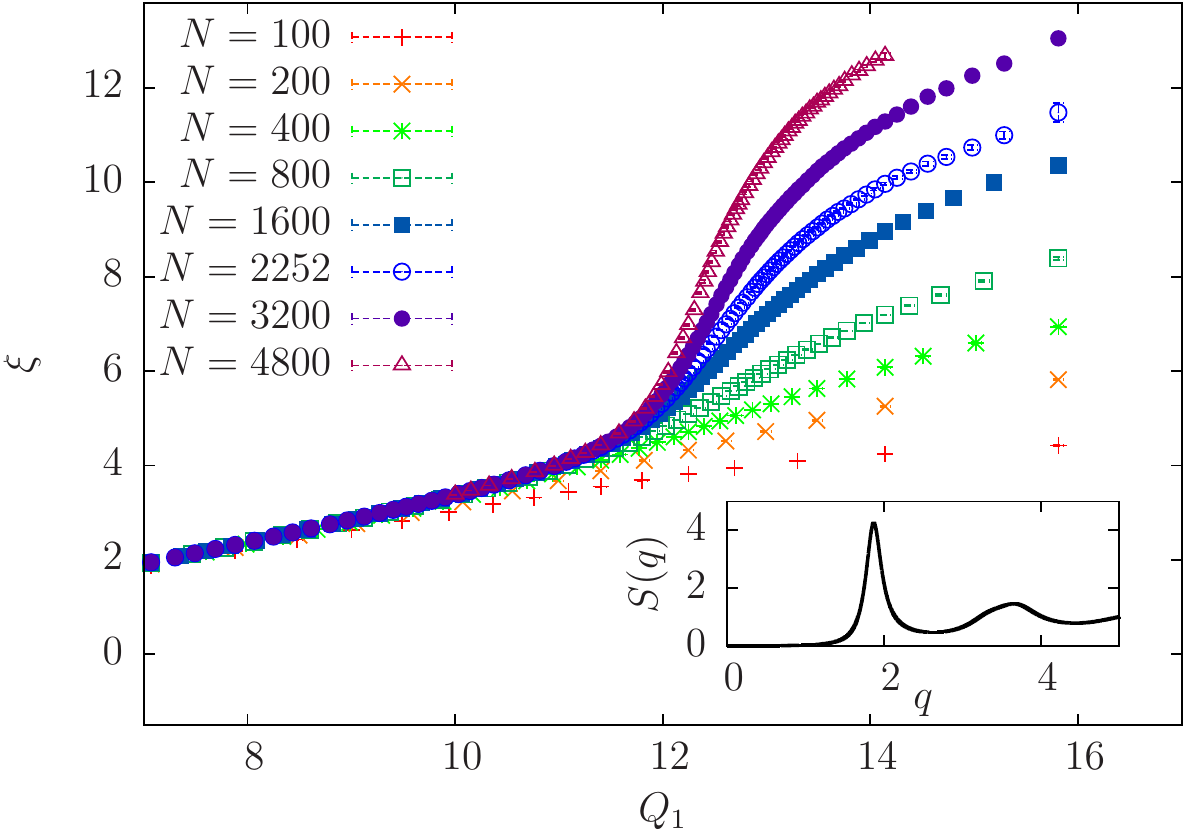}
  \caption{(Color online) Correlation length $\xi$ as a function of $Q_1$ for 8 different system sizes in the
   range $100 \leq N \leq4800$. The inset is a plot of $S(q)$ for the specific
    case when $N = 3200$ and $Q_1 = \sqrt{120}$.}
  \label{Fig:Correlation_length}
\end{figure}

The inset of Fig.~\ref{Fig:Correlation_length} shows a plot of $S(q)$. We assume that the correlation length $\xi$ is inversely proportional to the
width of the first peak in $S(q)$, and may thus be determined by a Lorentzian fit. The procedure is identical to that used in Ref.~\onlinecite{Moore_PRL_1999}, and the result is given in Fig.~\ref{Fig:Correlation_length}. For small values of $Q_1$ our results are
similar to Fig.~2 in Ref.~\onlinecite{Moore_PRL_1999}. However, when $Q_1 \approx 12$, a value that corresponds well with the critical coupling
of the freezing transition, we find a kink developing with increasing $N$, that clearly violates $\xi \propto Q_1$. This kink is not seen in
Fig.~2 of Ref.~\onlinecite{Moore_PRL_1999}. However, we note that the markers of that figure exhibits large scattering. Moreover, the authors did 
not consider larger values of $Q_1$.

\begin{figure}[tbp]
  \includegraphics[width=\columnwidth]{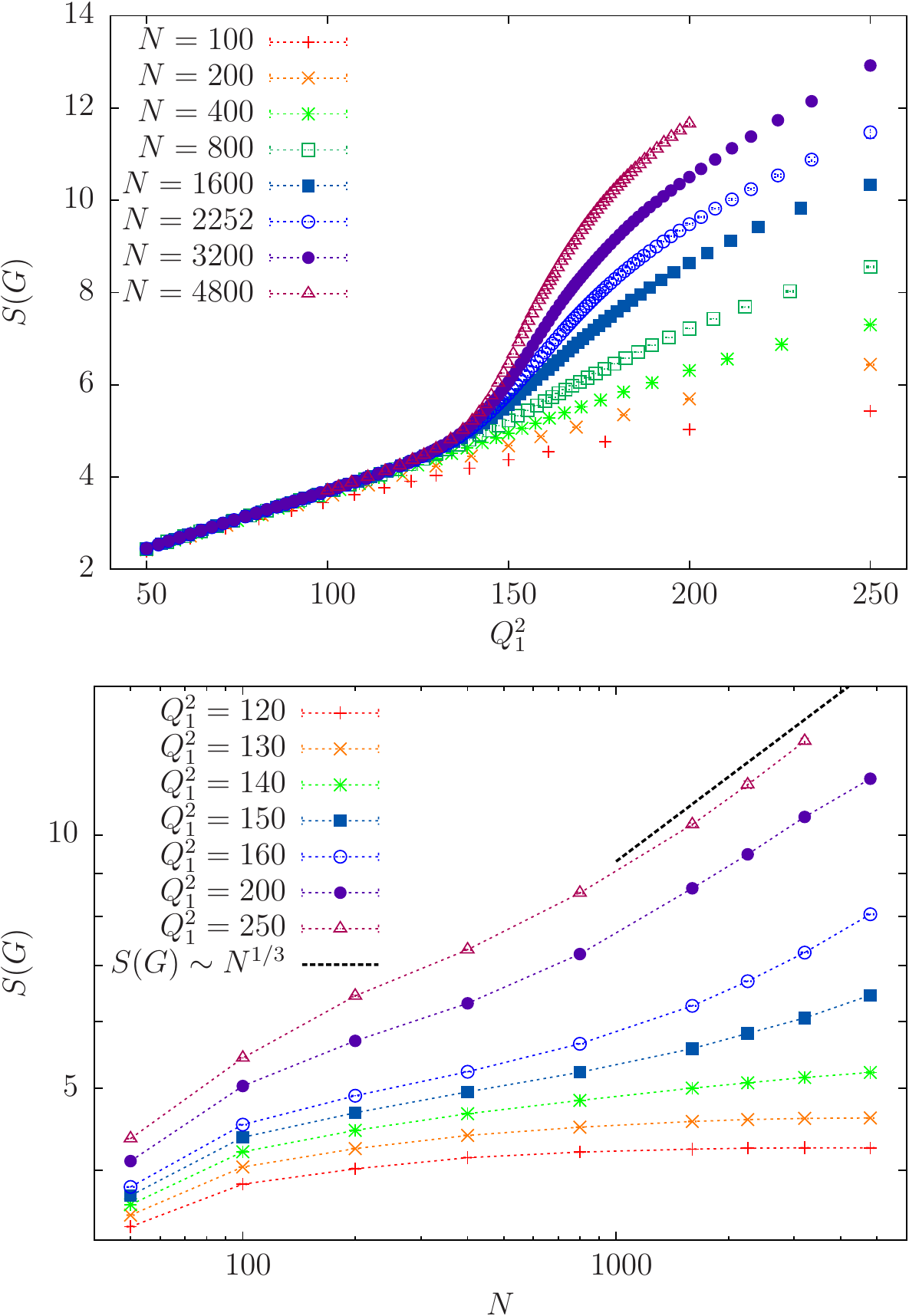}
  \caption{(Color online) Finite-size behavior of the peak value of the structure factor, $S(G)$, as a function of coupling $Q_1^2$ and
   size $N$. The upper panel shows $S(G)$ as a function of $Q_1^2$ for eight different sizes in the range $100 \leq N \leq 4800$. The
   lower panel is a log-log plot of $S(G)$ as a function of $N$ for seven fixed values of $Q_1^2$. The dashed line is a reference line
   that yields the expected finite-size behavior at the melting point according to KTHNY theory. Lines are guide to the eyes.}
  \label{Fig:Finite_size_behavior}
\end{figure}

A hallmark of a 2D solid is that translational correlations have a power-law decay,
$\langle \mathrm{e}^{i\bfG(\bfr-\bfr')} \rangle \sim |\bfr - \bfr'|^{-\eta_{\bfG}}$, where $\bfG$ is the reciprocal lattice vector, $\bfr$, $\bfr'$
are lattice points in the 2D solid and $\eta_{\bfG}$ is a temperature dependent exponent~\cite{Nelson_PRB_1979, Imry_PRB_1971}. Consequently, the first-order Bragg peak in $S(\bfq)$ will, as seen by Eq.~\eqref{Eq:Structure_function}, scale as $S(\bfG) \sim L^{2-\eta_{\bfG}}$, where $L \propto N^{1/2}$ is the spatial linear extent of the system. Now, by integrating
over the Bragg peak of a 2D solid~\cite{Dutta_PRL_1981}, the finite-size scaling of the azimuthally averaged first peak in $S(q)$, is given by
\begin{equation}
S(G) \sim L^{1-\eta_{\bfG}} \sim N^{(1-\eta_{\bfG})/2}.
\label{Eq:Peak_finite_size_scaling}
\end{equation}
Fig.~\ref{Fig:Finite_size_behavior} shows the
results for the maximum value of the first peak in $S(q)$ for a wide range of system sizes and for different values
of $Q_1^2$. As for $\xi$, we find that the peak value also exhibits a kink at $Q_1^2 \approx 140$ that should be associated
with an abrupt change in the translational correlations in the plasma. Indeed, when studying the finite-size behavior
more closely in the lower panel of Fig.~\ref{Fig:Finite_size_behavior}, the results show that when $Q_1^2 \leq 130$,
$S(G) \sim \text{const.}$ when $N$ increases. This is the behavior expected in the liquid phase, with exponentially
decaying translational correlations where $S(G)\sim \xi^2$. However, when $Q_1^2 \geq 150$, the results clearly show
that there is a positive slope that develops with increasing $N$, thus confirming the finite-size behavior of the 2D solid
given in Eq.~\eqref{Eq:Peak_finite_size_scaling}. When $Q_1^2 = 140$, it is difficult to determine whether the system is in
the solid phase or not, suggesting that $Q_1^2 = 140$ is close to the melting point of the OCP. Note that in
Fig.~\ref{Fig:Finite_size_behavior}, the height of the first-order peak in $S(q)$, $S(G) \approx 5$ when $Q_1^2 = 140$. 
This is consistent with the 2D freezing criterion for a crystal with long-range interactions (characterized by a divergent bulk modulus~\cite{DSFisher_1982})\cite{Caillol_JStatPhys_1982, Wang_JChemPhys_2011}.

A key prediction of the KTHNY theory is that $\eta_{\bfG} \leq 1/3$ in the solid phase, where the limiting value of $1/3$ is 
reached at the critical point of melting from a triangular lattice to the hexatic phase~\cite{Nelson_PRB_1979, DSFisher_1982, Franz_PRL_1994}.
As a result, in this scenario, $S(G)$ grows more rapidly with $N$ than $N^{1/3}$ for all $Q_1^2$ greater than the critical 
value; $S(G)$ grows as $N^{1/3}$ at the transition point; and $S(G)$ saturates in the liquid phase. Meanwhile, if the transition
were first-order, the limiting value of $\eta_{\bfG}$ would be smaller than $1/3$, so that $S(G)$ would grow more rapidly than 
$N^{1/3}$ at the transition point, i.e., the slowest possible growth of $S(G)$ in the crystalline phase would be faster than 
$N^{1/3}$. Consequently, we expect the slope of $\ln S(G)$ vs. $\ln N$ to be steeper than $1/3$ for all $Q_1^2$ in the crystalline
phase or, by the results above, for all ${Q_1^2}>140$. By determining the slope of $\ln S(G)$ vs. $\ln N$ at ${Q_1^2}\approx 140$, 
we could then determine if the transition is of KTHNY type or is first-order. However, as may be seen in Fig.~\ref{Fig:Finite_size_behavior}, 
the slopes of $\ln S(G)$ vs. $\ln N$ in the putative crystalline phase are {\it not} steeper than $1/3$ in our simulations.
However, the slopes steepen with increasing $N$, possibly converging towards the expected behavior in the thermodynamic limit. 
Therefore, we are unable to determine which type of transition occurs, nor whether melting proceeds via an intermediate 
hexatic phase (see below).

It is worth emphasizing that for ${Q_1^2}<140$, $S(G)$ appears to saturate to a finite value, as expected in a liquid, while, 
for ${Q_1^2}>140$, $S(G)$ does not appear to saturate, as expected in a crystal (although, as noted above, it does not grow as 
rapidly as expected). Therefore, the lower panel of Fig.~\ref{Fig:Finite_size_behavior} is also qualitatively consistent with a 
crystalline phase of the 2D OCP, which melts at ${Q_1^2}\approx 140$. Taken together with Fig.~\ref{Fig:Correlation_length} and 
the upper panel of Fig.~\ref{Fig:Finite_size_behavior}, this provides clear evidence for the existence of a low-temperature 
crystalline state of the OCP on a sphere, in agreement with previous 
studies~\cite{de_Leeuw_PhysicaA_1982, Caillol_JStatPhys_1982, Choquard_PRL_1983, Franz_PRL_1994}. Our results
contradict the claims made in Ref.~\onlinecite{Moore_PRL_1999} for the non-existence of a crystalline phase. 
The present work considers larger values of $Q_1$ and provides considerably better statistics than Ref.~\onlinecite{Moore_PRL_1999}.

The transition from the hexatic to the isotropic liquid phase is governed by fluctuations in the orientational order
parameter. These can be computed by the bond-orientational susceptibility which exhibits a peak at this transition. When this peak is 
obtained at a different point than the onset of quasi-long-range translational order, it indicates the existence of an intermediate hexatic
phase in between the crystalline and isotropic liquid phase~\cite{Qi_J_Chem_Phys_2010, Dillmann_J_Phys_2012}. 
On a sphere, we compute the bond-orientational susceptibility 
\begin{equation}
  \chi_6 = N\left(\left\langle|\Psi_6^2|\right\rangle-\left\langle\sqrt{|\Psi_6^2|}\right\rangle^2\right).
  \label{Eq:Bond-orientational_susceptibility}
\end{equation}
Here, we have
\begin{equation}
  \Psi_6^2 = \frac{1}{N^2}\sum_{i,j=1}^N \mathrm{e}^{\im 6 (\theta_{j|i}-\theta_{i|j})} \tilde{\psi}_{6,i}^* \tilde{\psi}_{6,j}.
  \label{Eq:Bond-orientational_susceptibility_I}
\end{equation}
Furthermore, the quantity
\begin{equation}
  \tilde{\psi}_{6,i} = \frac{1}{n_i}\sum_{a=1}^{n_i}\mathrm{e}^{\im 6\phi_{ia}}
  \label{Eq:Bond-orientational_susceptibility_II}
\end{equation}
may be thought of as the local bond-orientational order parameter of
particle $i$, where all bond angles $\phi_{ia}$ are measured with respect to the
closest nearest neighbor in the tangential plane of particle $i$. The
sum is over all $n_i$ nearest neighbors as determined by Voronoi
construction. In Eq.~\eqref{Eq:Bond-orientational_susceptibility_I},
the angle $\theta_{i|j}$ is the bond angle of the bond to the closest
nearest neighbor of particle $i$, measured with respect to the $i,j$
chord in the tangential plane of particle $i$. Thus, the chord line
combining the two particles for every term in the
sum of Eq.~\eqref{Eq:Bond-orientational_susceptibility_I}, serves as a 
line of reference for bond-orientational
order\cite{Prestipino_Giarritta_Physica_A_1992, Voogd_PhD_1998}.

\begin{figure}[tbp]
  \includegraphics[width=\columnwidth]{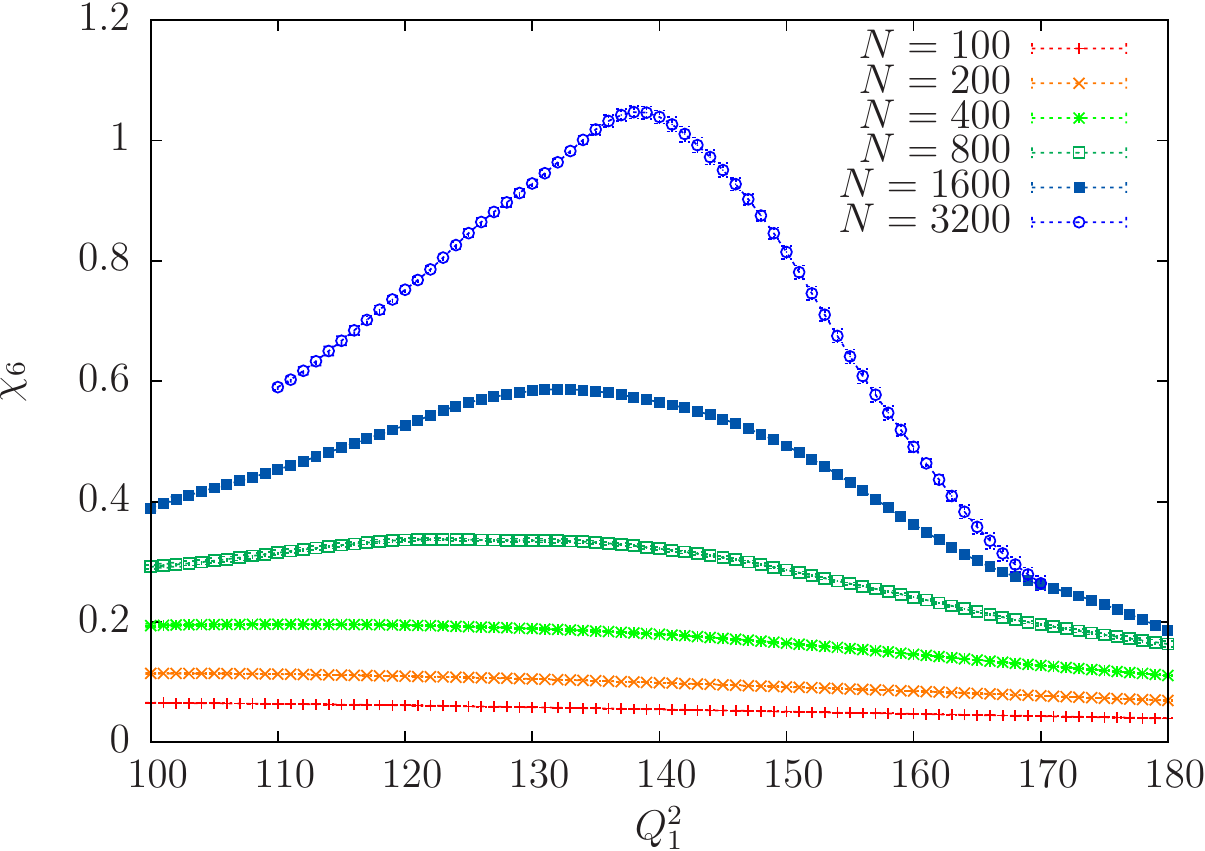}
  \caption{(Color online) Bond-orientational susceptibility $\chi_6$ as a function of $Q_1$ for 7 different system sizes in the
   range $100 \leq N \leq3200$. The position of the peak appears to converge to $Q_1^2 \approx 140$.}
  \label{Fig:Bond-orientational_suscetibility}
\end{figure}

In Fig.~\ref{Fig:Bond-orientational_suscetibility}, the results for $\chi_6$ are given. When the system size 
is large ($N \geq 800$), a peak is found. However, the value of $Q_1$ at which the peak occurs, appears to 
converge towards $Q_1^2 \approx 140$. Thus, with the resolution available, we
cannot confirm the existence of a hexatic phase since the position of the peak in $\chi_6$ does not
appear to converge to a coupling significantly different from $Q_1^2 \approx 140$. Our findings are consistent
with earlier works, which also have found no traces of an intermediate hexatic phase in these systems \cite{de_Leeuw_PhysicaA_1982, Caillol_JStatPhys_1982, Choquard_PRL_1983, Franz_PRL_1994}. 

\section{Results for an Unconventional Two-Component Plasma}

We now turn our attention to the full model in Eq.~\eqref{Eq:Model_S}, i.e., when both $Q_1$ and $Q_2$ are nonzero. In particular, we
consider how the translational ordering of both $w$ and $z$ particles is affected as we increase the coupling constant in the second
interaction-channel, $Q_2$. As for the OCP, we study translational correlations by measuring the structure factors $S_w(q)$, $S_z(q)$ defined by
Eq.~\eqref{Eq:Structure_factor} with $S(q)$, $g(R\theta) \rightarrow S_{w/z}(q)$, $g_{w/z}(R\theta)$. In addition, we also measure the
inverse dielectric constant for charges with interaction of type 2, given by
\begin{equation}
\epsilon_{22}^{-1} = 1 - \frac{\pi Q_2^2 R^2}{A} \left \langle \left(\sum_{i=1}^{N}\hat{\bfz}_{i} - \sum_{a=1}^{N}\hat{\bfw}_{a}\right)^2 \right \rangle.
\label{Eq:Inverse_dielectric_constant}
\end{equation}
This quantity measures the screening properties for charges interacting with $Q_2$, and it signals a charge-unbinding transition involving $z$-
and $w$-particles~\cite{Herland_PRB_2012}.

\begin{figure}[tbp]
\includegraphics[width=\columnwidth]{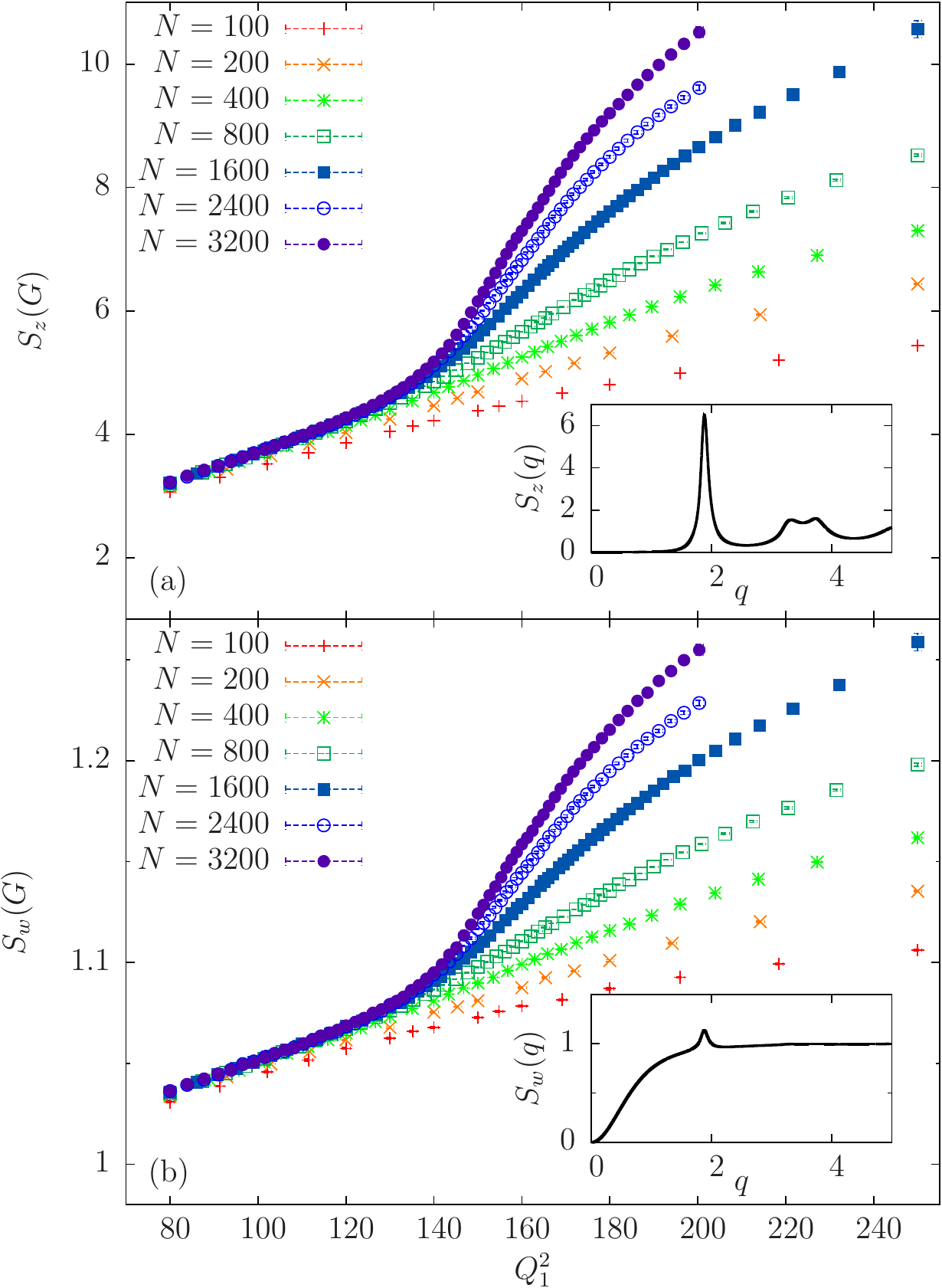}
\caption{(Color online) Results from the Monte Carlo simulations when $Q_2^2 = 1$ and $\eta = 10^{-3}$. Panel (a) shows the height of the
first-order peak in the structure factor for the $z$ particles, $S_z(G)$, as a function of the coupling $Q_1^2$ for seven different sizes in
the range $100 \leq N \leq 3200$. Panel (b) shows the height of the first-order peak in the structure factor for the $w$ particles, $S_w(G)$,
as a function of the coupling $Q_1^2$ for seven different sizes in the range $100 \leq N \leq 3200$. In order to give an impression on how a
typical structure factor looks like, the insets of panel (a) and (b) show plots of $S_{z}(q)$ and $S_{w}(q)$ for the specific case when
$N = 800$, $Q_1^2 = 180$ and $Q_2^2 = 1$.}
  \label{Fig:Combined_panels}
\end{figure}

\begin{figure}[tbp]
  \includegraphics[width=\columnwidth]{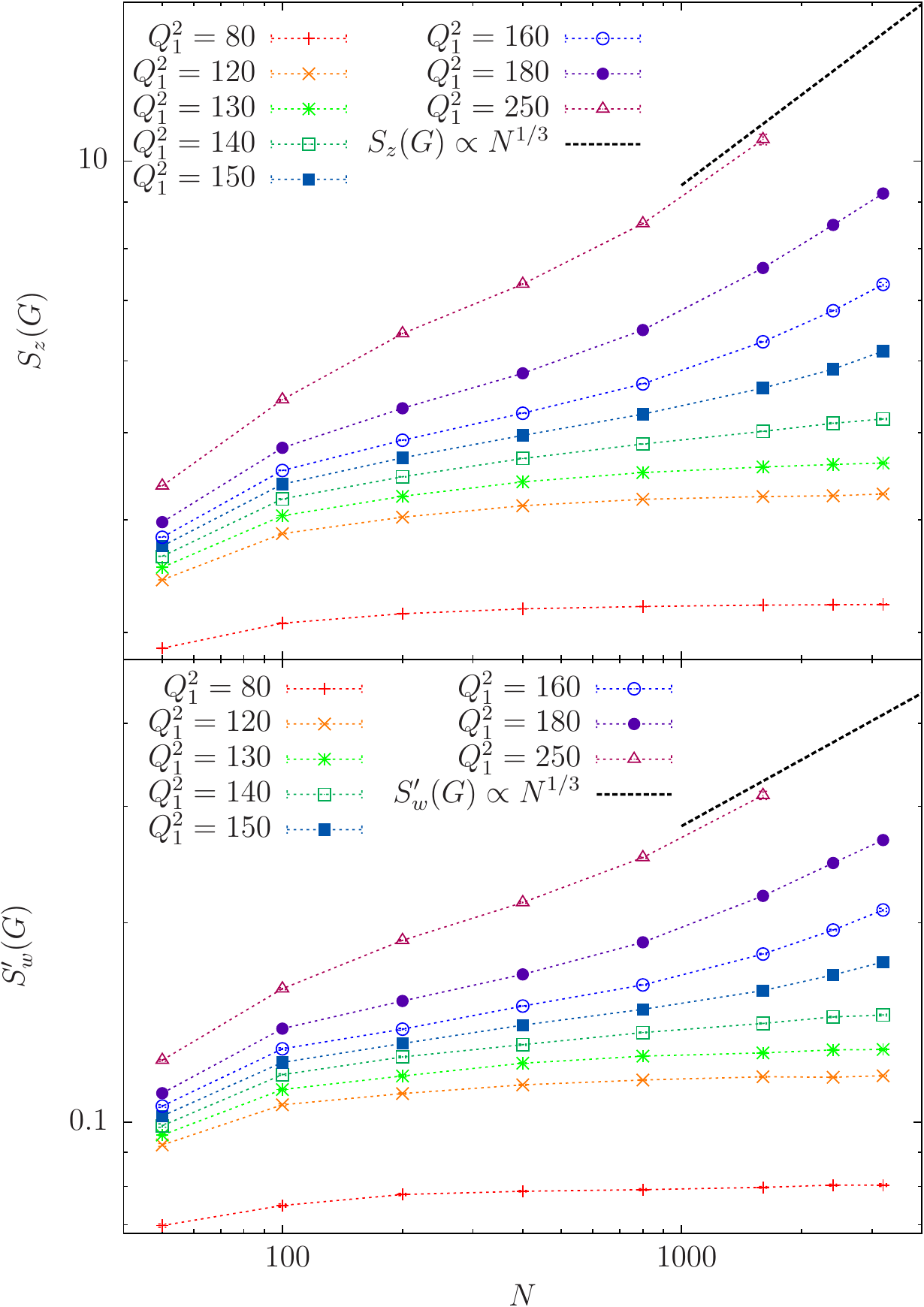}
  \caption{(Color online) Log-log plots of the results in panel (a) and (b) in Fig.~\ref{Fig:Combined_panels}. Both panels show
    the height of the first-order peak in the structure factor as a function of size $N$. In the lower panel, $S_w^{\prime}(G)$ is
    the height of the peak of $S_w(G)$ when we have subtracted the regular part in order to properly extract the singular finite-size
    behavior of $S_w(q)$ in a log-log plot. The solid lines are reference lines that yield the expected finite-size behavior at the
    melting point according to KTHNY theory. Lines are guide to the eyes.}
  \label{Fig:Structure_factor_both_comp}
\end{figure}

\begin{figure}[tbp]
  \includegraphics[width=\columnwidth]{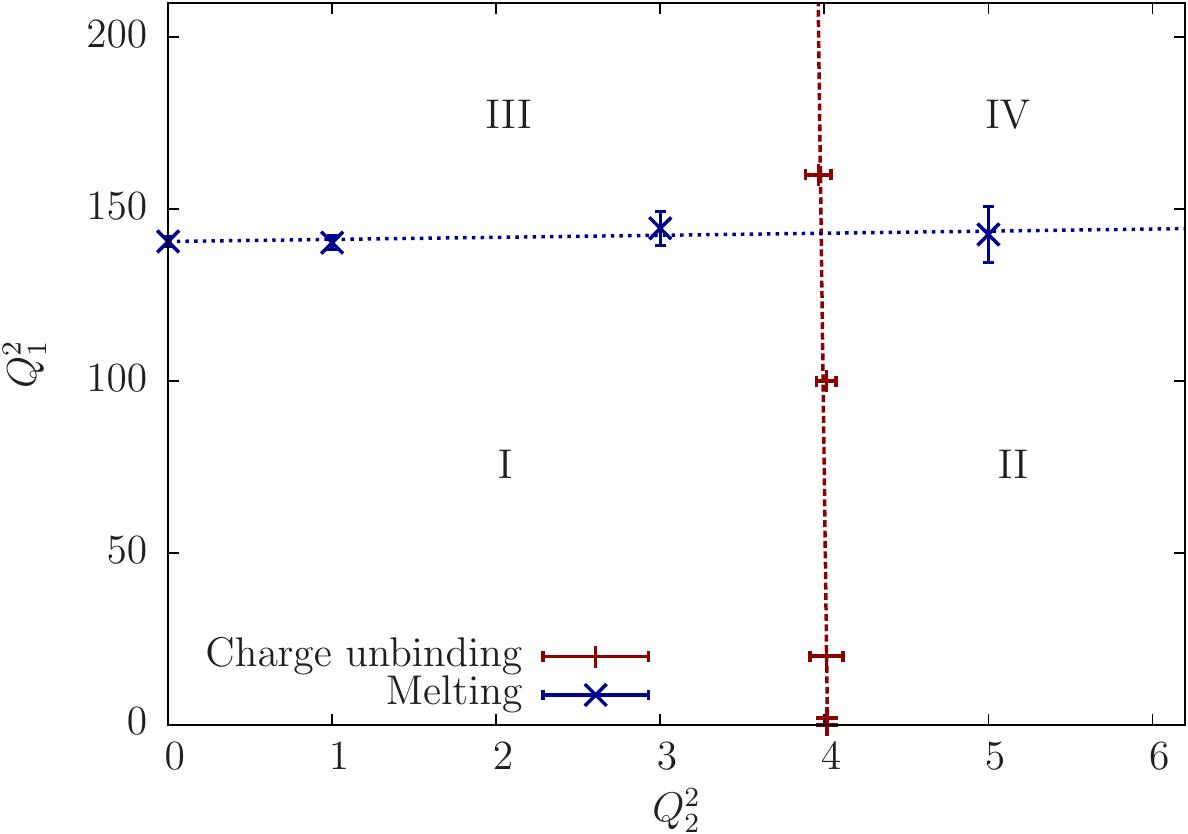}
  \caption{(Color online) The phase diagram as a function of $Q_1^2$ and $Q_2^2$. The
    dashed red line is
  the charge unbinding transition, at which $z$ and $w$ particles become bound together for
  at $Q_2^2$ above this line. 
  The dotted blue line is the melting line of the Wigner crystal.
  In Phases I and II, the $z$ particles are in a liquid state; in Phases III and IV, the $z$ particles form
 a Wigner crystal. In Phases I and III, the $w$ and $z$ particles are
 unbound; in Phases II and IV, they are bound into molecules comprised of one $z$
 and one $w$ particle. See the text for details.}
  \label{Fig:plot_phase_diagram}
\end{figure}

In Fig.~\ref{Fig:Combined_panels}, results are given for the height of the first-order peak in the structure factor for
component $z$ and $w$, for the case when $Q_2^2 = 1$. Apart from the fact that the height of the peak in the structure factor is much larger for the $z$ 
particles than the $w$-particles, the size- and  $Q_1^2$-dependence of the peaks are qualitatively very similar  for the two components. In particular, 
they both exhibit a kink at $Q_1^2 \approx 140$, which should be associated with melting of a 2D solid, similar to the OCP case in the upper panel of 
Fig.~\ref{Fig:Finite_size_behavior}. Specifically, when we extract the finite-size behavior in the log-log plots in Fig.~\ref{Fig:Structure_factor_both_comp}, 
we find that both components exhibit $S(G) \sim \text{const.}$, consistent with being in the liquid phase, when $Q_1^2 \leq 130$. When $Q_1^2 \geq 150$, 
the results clearly show that there is a power-law dependence on  $N$, consistent with the finite-size behavior of a 2D solid. These results are consistent 
with the phase diagram in Fig.~\ref{Fig:plot_phase_diagram}.

The inverse dielectric constant $\epsilon_{22}^{-1}$ is measured to be zero to the left of the red
line in Fig.~\ref{Fig:plot_phase_diagram}. Thus, the $w$ and $z$ particles are in a metallic state
regardless of the change in the structural properties when $Q_1^2 \approx 140$. This is most salient 
with respect to the second type of Coulomb interaction (which has an effective strength that is determined 
by $\epsilon_{22}^{-1}$). In the liquid phase, it is clear that $w$ and $z$ particles are in a metallic state.
In the crystalline phase, there are interstitials and vacancies in the crystal, so that a finite fraction 
of $w$ and $z$ particles should be considered as unbound particles that are able to screen test particles 
interacting with charges of type 2, thereby leading to  $\epsilon_{22}^{-1}=0$. At larger values of $Q_2^2$,
there is a transition at which $w$ and $z$ particles are bound into molecules. For $Q_2^2$ above this transition 
point, which is at $Q_{2,c}^2 \approx 4$, $\epsilon_{22}^{-1}$ has a non-zero value, as found in
Ref.~\onlinecite{Herland_PRB_2012}. Although the $w$ particles are able to screen the type 2 interaction
when $z$ particles form a Wigner crystal, their translational correlations exhibit signatures of a 2D solid 
(as seen in Fig~\ref{Fig:Combined_panels}), attributed to a higher probability of the $w$ particles to be 
co-centered with $z$ particles due to the attractive inter-component interactions of type 2. On average, a 
finite fraction of the $w$ particles should be considered as bound to the $z$ particles, thus adapting to the 
2D crystalline structure that is created by the strong repulsive interactions among the $z$ particles, when 
$Q_1^2 > 140$. The signatures of freezing of the $w$-particles is thus an effect which is induced by the freezing 
of the $z$-particles.

Hence, we can summarize the situation as follows, as depicted
in Fig.~\ref{Fig:plot_phase_diagram}. In Phase I, the $w$ and $z$ particles are unbound and are separately in a liquid state. In 
Phase II, the $w$ and $z$ particles are bound into molecules that are neutral with respect to the second type of Coulomb interaction,
and these molecules form a liquid. In Phase III, the $w$ and $z$ particles are unbound;
the $z$ particles form a Wigner crystal while the $w$ particles form a liquid, albeit one
with modulated density due to its interaction with the Wigner crystal.
In Phase IV, the $w$ and $z$ particles are bound into molecules forming a Wigner crystal.
In Appendix~\ref{sec:transitions},
we explain the details of how the transition lines were obtained.

We consider phase III to be a 2D counterpart of the situation that was reported for a three-dimensional system in Ref.~\onlinecite{Dahl_PRL_2008}.
This work considered a two-component rotating BEC with a negative dissipationless Andreev-Bashkin drag~\cite{Andreev_JETP_1975}.
It was found that in this mixture, a situation may arise where the component with the smallest stiffness will be a
\textit{modulated vortex liquid}. That is, the soft component breaks translational symmetry while exhibiting an unbroken symmetry in
order parameter space. The vortices of the soft component are likely to be co-centered with the vortices of the stiffest component, and will
thus adapt to the spatial structure of the latter. As shown in
Ref.~\onlinecite{Herland_PRB_2012}, the MR plasma corresponds to a 2D
two-component rotating BEC with negative drag,
where the $z$ component is stiffer than the $w$ component when $Q_1^2 > 0$.

\section{Summary and Conclusions}

In summary, we have considered the melting of an unconventional 2D two-component plasma on a sphere with particles interacting in two
different channels, which may be viewed as an analogous plasma describing a non-Abelian Ising-type quantum Hall state or a realization 
of a two-component two-dimensional Bose-Einstein condensate with inter-component non-dissipative drag. In the limiting case where there 
are no interactions of type 2 ($Q_2=0$), the system is a standard 2D one-component plasma. Both for the one-component plasma 
and the unconventional two-component plasma, we find that the system freezes on a sphere for large enough inter-particle interactions. For 
the two-component plasma, the $w$-particles do not have strong intra-component interactions, but still show signatures of forming a 2D solid.
This is attributed to the attractive inter-component interactions with the $z$-particles that leads to a higher probability of the $w$-particles 
to be co-centered with $z$-particles. The $w$-particles nonetheless form a  metallic state. We have also examined the possible existence of an 
intermediate hexatic phase in the one-component plasma. Our results show that the value of $Q_1$ where diclinations unbind and orientational 
order is lost, cannot be distinguished from the value of $Q_1$ where dislocations unbind and quasi-long-range translational order is lost.  

\acknowledgments
We acknowledge useful discussions with J.~S. H\o ye, and I.~B. Sperstad. E.~B., P.~B., C.~N. and A.~S. thank the Aspen Center for Physics for hospitality 
and support under the NSF Grant No.~$1066293$. E.~V.~H. thanks NTNU for financial support. E.~B. was supported by Knut and Alice Wallenberg Foundation 
through the Royal Swedish Academy of Sciences Fellowship, Swedish Research Council and by the National Science Foundation CAREER Award No. DMR-0955902.
V.~G. was supported by NSF Grant No. PHY-0904017. C.~N. was supported in part by the DARPA QuEST program. L.~R. acknowledges support by the NSF through DMR-1001240 and MRSEC DMR-0820579. A.~S. was supported by the Norwegian Research Council under Grants No. 205591/V30 and 216700/F20. The work was also supported through the Norwegian consortium for high-performance computing (NOTUR).

\appendix
\section{Determination of the transition lines}
\label{sec:transitions}

We now discuss the determination of the phase transition lines
in Fig.~\ref{Fig:plot_phase_diagram} in more detail.

First, consider the line (red in Fig.~\ref{Fig:plot_phase_diagram}) at which the
$z$ and $w$ particles unbind. In this work, we find the critical point
of the BKT transition by curve-fitting the inverse dielectric constant to a
logarithmic finite-size scaling relation with one free parameter (see
Appendix C in Ref.~\onlinecite{Herland_PRB_2012}). This means that we
assume that the transition is a BKT transition as we
use the BKT value of the universal jump in the finite-size scaling
relation. Thus, it is a slightly less self-consistent approach than
what was used in Ref.~\onlinecite{Herland_PRB_2012}, but still, one can
regard this as a verification of the BKT nature, as one should not
expect a good fit to the scaling relation if the transition is of a
different nature.\footnote{Using two free parameters, as we did
in Ref.~\onlinecite{Herland_PRB_2012}, is highly demanding in terms of
statistical quality. A one-parameter fit requires much less statistics.}

\begin{figure}[t!]
  \includegraphics[width=\columnwidth]{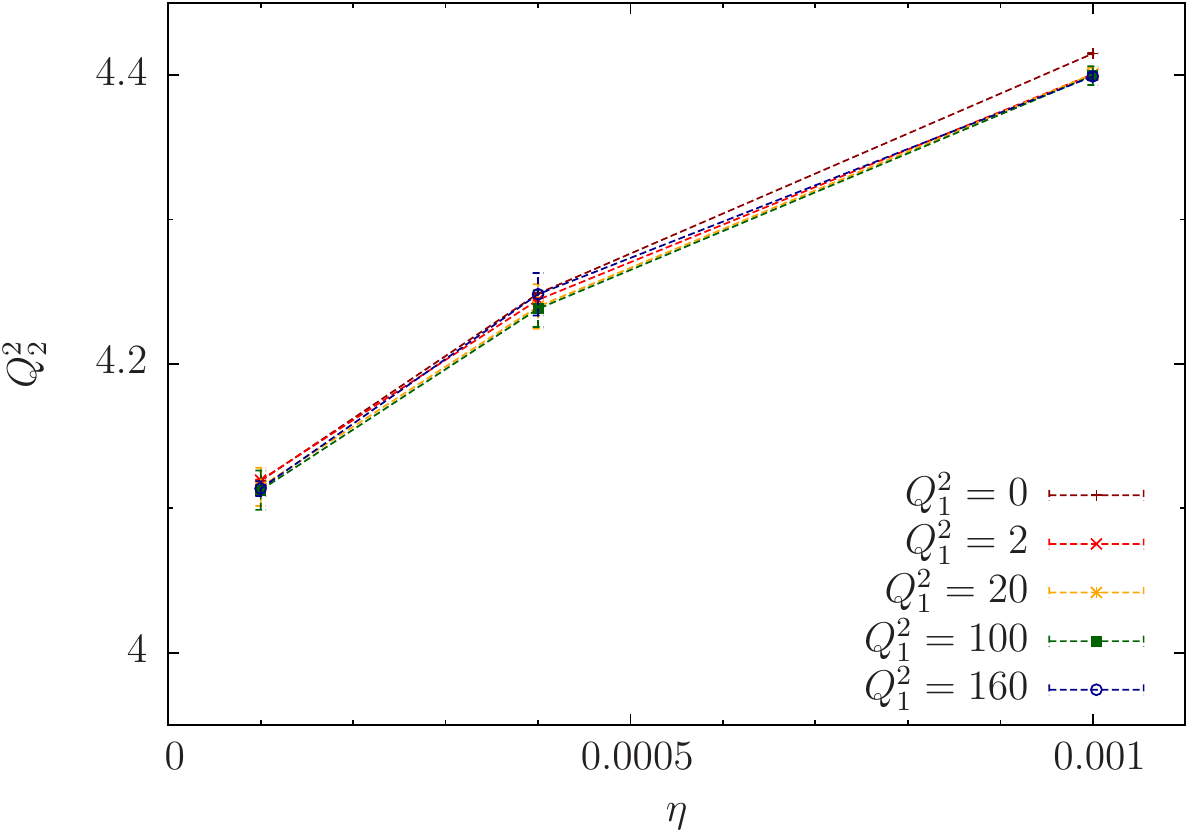}
  \caption{(Color online) The transition point is determined for $\eta\rightarrow 0$ by extrapolating
  the $Q_{2,\text{c}}^2$ values obtained for non-zero $\eta$, assuming
  a power-law dependence. Lines are guide to the eyes.}
  \label{Fig:plot_QQ_sfa_eta}
\end{figure}

The curve-fitting was performed according to the description in
Appendix C in Ref.~\onlinecite{Herland_PRB_2012} for sizes $N = 70$,
$100$, $150$, $200$, $300$, and $500$, for $Q_1^2 = 20$, $100$, and
$160$ and for densities $\eta = 0.001$, $0.0004$, and $0.0001$. In
Fig.~\ref{Fig:plot_QQ_sfa_eta}, the results for the
transition point $Q_{2,\text{c}}^2$ as a function of $\eta$ and
$Q_1^2$ are given. We have also included the
results for $Q_2^2 = 0$ and $2$ from Ref.~\onlinecite{Herland_PRB_2012}.
In order to obtain a crude estimate of the transition temperature in
the low density limit, we extrapolate to $\eta = 0$ by fitting the results for finite $\eta$ to a power law $Q_{2,\text{c}}^2(\eta) = Q_{2,\text{c}}^2 + a\eta^b$, where $Q_{2,\text{c}}^2$, $a$, and $b$ are free parameters.
The estimates we find are:
\begin{equation}
\begin{array}{llll}
Q_{2,\text{c}}^2 = 4.016 \pm 0.002 & & \text{ for } & Q_1^2 = 0 \\
Q_{2,\text{c}}^2 = 4.015 \pm 0.004 & & \text{ for } & Q_1^2 = 2 \\
Q_{2,\text{c}}^2 = 4.013 \pm 0.101 & & \text{ for } & Q_1^2 = 20 \\
Q_{2,\text{c}}^2 = 4.012 \pm 0.060 & & \text{ for } & Q_1^2 = 100 \\
Q_{2,\text{c}}^2 = 3.963 \pm 0.070 & & \text{ for } & Q_1^2 = 160
\end{array}
\end{equation}
These values are plotted in the phase diagram in Fig.~\ref{Fig:plot_phase_diagram},
and we take the phase boundary to be the best fit straight line running through them.

\begin{figure}[t!]
  \includegraphics[width=\columnwidth]{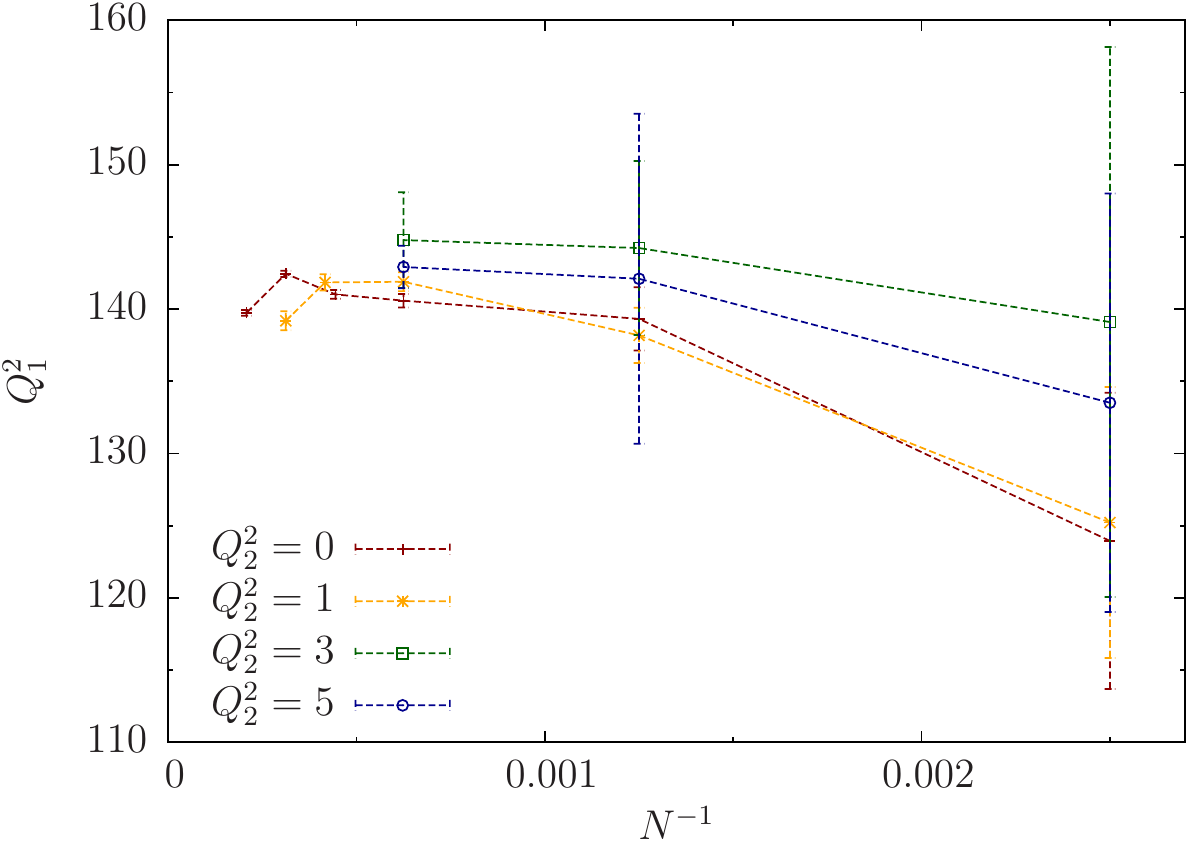}
  \caption{(Color online) The transition point $Q_{1,\text{c}}^2$ as a function of system size for
  $Q_2^2 = 0$, $1$, $3$, and $5$, estimated by the maximum of the second
  derivative of $S_z(G)$. See text for details. Lines are guide to the eyes.}
  \label{Fig:plot_melting}
\end{figure}

We now consider the Wigner crystal melting transition, depicted by the blue line
in Fig.~\ref{Fig:plot_phase_diagram}.
This transition is found by measuring the value of $Q_1^2$ at which
$S_z(G)$ attains its maximum second-derivative.
(See Fig.~\ref {Fig:Finite_size_behavior} and Fig.~\ref{Fig:Combined_panels}
in the paper for example.) 
In Fig.~\ref{Fig:plot_melting}, we show the estimates of the transition
point $Q_{1,\text{c}}^2$ as a function of inverse system size $N^{-1}$
for $Q_2^2 = 0$, $1$, $3$, and $5$. 
The transition points are estimated by averaging the results for $N \geq 800$, with errors determined by a bootstrap analysis. The estimates we find are:
\begin{equation}
\begin{array}{llll}
Q_{1,\text{c}}^2 = 140.6 \pm 1.5 & & \text{ for } & Q_2^2 = 0 \\
Q_{1,\text{c}}^2 = 140.3 \pm 2.0 & & \text{ for } & Q_2^2 = 1 \\
Q_{1,\text{c}}^2 = 144.4 \pm 4.9 & & \text{ for } & Q_2^2 = 3 \\
Q_{1,\text{c}}^2 = 142.6 \pm 8.2 & & \text{ for } & Q_2^2 = 5
\end{array}
\end{equation}
The phase diagram in Fig.~\ref{Fig:plot_phase_diagram} is obtained by
using these values, and we take the phase boundary to be the best fit straight line running through them.

\end{document}